\documentclass[a4paper]{article}
\usepackage[psamsfonts]{amssymb}
\usepackage{amsmath}
\usepackage{epsfig}\author{H. Mohseni Sadjadi\footnote{mohseni@phymail.ut.ac.ir}
\\ {\small School of Physics, College of Science,  University of Tehran,}
\\ {\small North Karegar Ave., Tehran, Iran.}}
\title{ The effect of geometry on charge confinement in three dimensions}
\begin{document}
\maketitle
\begin{abstract}
We consider three dimensional BTZ black-hole (covering space) metric and show
that in contrast to the flat case the Maxwell theory is not
confining. We also study the effect of the curvature on screening
behavior of Maxwell-Chern-Simons model in this space-time.


\end{abstract}
\section{Introduction}
It has been proposed that the infrared behavior of quantum
chromodynamics, $QCD$, may be responsible for the confinement of
quarks and gluons. But the concept of quark confinement, i.e. the
increase of potential between quarks with separation, could not be
verified to be true using the perturbation methods (because of
infrared singularities) and must be studied nonperturbatively.
Because of computational difficulties, these kinds of calculations
can be done in an equivalent lower dimensional model to obtain
clues to the four dimensional space-time.

In two dimensional quantum electrodynamics ($QED_2$), known as
Schwinger model \cite{sc}, when dynamical fermions are massless,
the system is in screening phase, i.e. external charges are
screened by dynamical fermions. In the massive $QED_2$, depending
on the charge of dynamical fermions, the system can be in
confining or screening phase \cite{col}. For $QCD_2$, in
\cite{gross} it has been shown that a heavy probe charge is
screened by dynamical massless fermions both in the case when the
source and the dynamical fermions belong to the same
representation of the gauge group and also in the case when the
representation of the probe charge is smaller than the
representation of the massless fermions.

The confining and screening aspects of gauge theories in three
dimensions has been the subject of several studies, e.g.,  see
\cite{poly}. In $(2+1)$ dimensions, in the absence of vacuum
polarization Maxwell theory is confining . In the presence of
massive dynamical fermions (corresponding to the vacuum
polarization), the behavior of the system becomes the same as
Maxwell-Chern-Simons theory, which is screening \cite{ab}.

A dielectric constant which is larger than unity results screening
effect. If instead one considers an ideal case in which the
dielectric constant of the vacuum vanishes then due to
antiscreening effect the energy of the system becomes infinite
unless we add another opposite test charge to the system. So
models that can mimic the role of such a non-trivial dielectric
medium, e.g. by inserting additional fields in the gauge theories,
have been proposed to study quark confinement \cite{rev}. In the
same way similarities between the behavior of the gauge fields in
an effective non--uniform dielectric medium and in a static
gravitational background \cite{mol} lead us to study the effects
of geometry and topology of space -time on confining structure of
gauge theories. In \cite{vol}, a scenario for a possible
correspondence between basic notions of gauge theory and gravity
was discussed. There was claimed that confinement in gauge theory
corresponds to black holes in gravity and one can try to describe
hadrons as strongly curved universes or black holes. In this
picture the horizon corresponds to the surface of the bag in a bag
model. In this way one can relate the mystery of curvature
singularities in classical general relativity with the mystery of
quark confinement.

In this work, using a specific example,  we try to elucidate how
the confining structure of gauge theories can be affected by
topology and geometry of a three dimensional space-time.

In $(1+1)$ dimensions the same problem has been studied in
\cite{eb},\cite{sad}, and it has been shown that the curvature of
the space-time, affects the confining and screening behavior of
$QED_2$. In \cite{sah} the effect of  nontrivial global properties
of a three dimensional space-time with conical singularity,
produced by a cosmic string or a point mass, on electromagnetic
field of a point charge has been discussed.

In this paper we consider Maxwell and  Maxwell- Chern-Simons
theories in an anti de Sitter space-time: upper half space
equipped with the Poincare metric in three dimensions. The
interaction potential of charges is obtained by solving the
classical equations of motion of the gauge fields. It is shown
that the curvature of the space-time, destroys the confining
aspect of Maxwell theory, encoded in the logarithmic behavior of
charges interaction potential. We find that the
Maxwell-Chern-Simons theory remains in screening phase like the
flat case. Using Dirac determinant on curved space-time, we
discuss the role of vacuum polarization in screening of test
charges. Indeed, fermion modes in $QED_3$,  will be  integrated
out to incorporate quantum (fermion loop) effects in our
subsequent classical analysis.

Note that the Poisson equation in a general curved space-time can
not be solved exactly, then we have restricted ourselves to a
specific curved space-time, in which the destruction of
(logarithmic) confining behavior of the potential, due to
geometrical effects, may be shown exactly.

Through the paper we use $\hbar=c=G=1$ units.

\section{Confining aspects of Maxwell theory on upper half space}

Consider the upper half-space $H^3=\{(t,x,y);y>0\}$, equipped with
the metric
\begin{equation}\label{1}
ds^2=\frac{l^2}{y^2}(dt^2-dx^2-dy^2).
\end{equation}
In terms of $l$, the scalar curvature is given by $R=-6/l^2$. This
space-time can be considered as the covering space of BTZ black
hole in the ground state \cite{ads}. The electrodynamics (Maxwell)
action in the presence of the current $J^{\mu}$ is
\begin{equation} \label{2}
S_{Max.}=\int_{H^3}\left(-\frac{1}{4}\sqrt{g}g^{\mu \nu}g^{\lambda
\beta}F_{\mu \lambda}F_{\nu
\beta}+\sqrt{g}J^{\mu}A_{\mu}\right)d^3x.
\end{equation}
$F_{\mu \nu}= \partial_{\mu}A_{\nu}-\partial_{\nu}A_{\mu}$ is the
field strength tensor, $g_{\mu \nu}$ are the metric tensor
components: $g_{00}=-g_{11}=-g_{22}=l^2/y^2$ and $\sqrt{g}=
l^3/y^3$ where $g$ is determinant of the metric. The classical
equation of motion of the gauge fields is
\begin{equation} \label{3}
\frac{1}{\sqrt{g}}\partial_{\nu}\sqrt{g}F^{\nu
\sigma}=-J^{\sigma}.
\end{equation}
We assume that the covariantly conserved current $J^{\mu}$ is
formed of two static opposite charges $e_1=e$ and $e_2=-e$,
located at $X_1=(t,x_1,y_1)$ and $X_2=(t,x_2,y_2)$ respectively
\cite{wein}:
\begin{eqnarray} \label{4}
J^{\mu}&=&\frac{1}{\sqrt{g}}\sum^2_{n=1}e_n\int
\delta^2(X-X_n)dt_n \\ \nonumber&=&
\frac{e}{\sqrt{g}}[\delta(x-x_1)\delta(y-y_1)-\delta(x-x_2)\delta(y-y_2)]\delta^{\mu
0}.
\end{eqnarray}
The poisson equation for the field $A_0$, is then
\begin{equation} \label{5}
\partial_{y}y\partial_{y}A_0+y\partial_x^2A_0=le[\delta(x-x_1)
\delta(y-y_1)-\delta(x-x_2)\delta(y-y_2)].
\end{equation}
The solution of this equation may be found by constructing the
Green's function
\begin{equation}\label{6}
\partial_{y}y\partial_{y}G(x,y;x',y')+y\partial_x^2G(x,y;x',y')
=l\delta(x,x')\delta(y,y')
\end{equation}
If we insert the Fourier expansion
\begin{equation} \label{7}
G(x,y;x',y')=\frac{1}{2\pi}\int_{-\infty}^{\infty}G_k(y,y')e^{ik(x-x')}dk,
\end{equation}
in the eq.(\ref{6}), we obtain
\begin{equation} \label{8}
\partial_yy\partial_yG_k(y,y')-yk^2G_k(y,y')=l\delta(y,y').
\end{equation}
In terms of the modified Bessel functions the solution of the
above equation is
\begin{equation} \label{9}
G_k(y,y')=-l I_0(ky_<)K_0(ky_>),
\end{equation}
where $y_>(y_<)$ is the larger (smaller) value of $y$ and $y'$.
This solution leads to well defined $A_0$ for both $y\to 0$ and
$y\to \infty$ limits. Inserting (\ref{9}) in (\ref{7}) results
\begin{equation} \label{10}
G(x,y;x',y')=\frac{-l}{\pi}\int_{0}^{\infty}
I_0(ky_<)K_0(ky_>)\cos\left(k(x-x')\right)dk.
\end{equation}
Using eq.(6.672.4) of \cite{table}, we obtain
\begin{equation} \label{11}
G(x,y;x',y')=-\frac{1}{2\pi}\frac{l}{\sqrt{y_>y_<}}
Q_{-\frac{1}{2}}\left[\frac{y_>^2+
y_<^2+(x-x')^2}{2y_>y_<}\right].
\end{equation}
$Q$ is the associated Legendre function. In the surface
$t=constant$, the geodesic distance of the points $(t,x,y)$ and
$(t,x',y')$ is
\begin{equation} \label{12}
L=l\cosh^{-1}\left[1+\frac{(y-y')^2+(x-x')^2}{2yy'}\right].
\end{equation}
Eq.(\ref{11}) can be rewritten as
\begin{equation} \label{13}
G(x,y;x',y')=\frac{-1}{2\pi}
\sqrt{\frac{l}{y}}\sqrt{\frac{l}{y'}}Q_{-\frac{1}{2}}\left(\cosh(\frac{L}{l})\right).
\end{equation}
Note that eq.(\ref{6}) is not invariant under scale transformation
($x\to \lambda x$, $y\to \lambda y$) which leaves the distance $L$
invariant, hence the potential, in contrast to the flat case,
depends on both the distance $L$ and the position of the charges.

The gauge field $A_0$, is then
\begin{equation} \label{14}
A_0(x,y)= e[G(x,y;x_1,y_1)-G (x,y;x_2,y_2)].
\end{equation}
Putting back eq.(\ref{14}) into eq.(\ref{2}) we obtain the energy
of the charges as
\begin{eqnarray} \label{15}
U&=&-\int \mathcal{L} \sqrt{g} dxdy=-\frac{1}{2}\int J^0A_0 \sqrt{g}dxdy\\
\nonumber
&=&-\frac{e^2}{2}[G(x_1,y_1;x_1,y_1)+G(x_2,y_2;x_2,y_2)-2G(x_1,y_1;x_2,y_2)].
\end{eqnarray}
$\mathcal{L}$ is Lagrangian density and the energy is measured by
an observer whose velocity is parallel to the global timelike
Killing vector of the space-time. \\
$-e_i^2/2[G(x_i,y_i;x_i,y_i)]$ is the self-energy of the charge
$i$, and \begin{eqnarray} \label{16}
U_{int.}&=&{e^2}[G(x_1,y_1;x_2,y_2)] \\
\nonumber &=&-\frac{e^2}{2\pi}
\sqrt{\frac{l}{y}}\sqrt{\frac{l}{y'}}Q_{-\frac{1}{2}}\left(\cosh(\frac{L}{l})\right),
\end{eqnarray}
is the interaction energy of the charges. To find the same
computation in the background of Schwarzschild black hole, based
on the total mass variation law of Carter \cite{cart}, see
\cite{smith}.

For widely separated charges $L\to \infty$, using \cite{table}
\begin{equation}\label{h}
 Q_{n-\frac{1}{2}}\left(cosh(z)\right)={\Gamma(n+\frac{1}{2})\over\Gamma(n+1)}\sqrt{\pi}
e^{-(n+\frac{1}{2})z}F(\frac{1}{2},n+\frac{1}{2},n+1,e^{-2z}),
\end{equation}
one can show that $Q_{-1/2}\left(\cosh(L/l)\right)$ decays as
$\cosh^{-1/2}\left(L/l\right)$ and the interaction energy tends to
$-0$ which shows that the system is not in confining phase.  In
the limit $l\to \infty$, and $g_{\mu \nu}=\eta_{\mu \nu}$, the
Green function (\ref{13}), by ignoring some coordinates
independent parameters (which is allowed due to the form of the
equation (\ref{6}), i.e. adding a constant to $G$ does not change
the eq.(\ref{6})), becomes
\begin{equation} \label{17}
G_{flat}=\frac{1}{2\pi}ln(L).
\end{equation}
We have used $\lim_{x\to 1}
Q_{-1/2}(x)=1/2\left(\ln(2)-\ln(x-1)\right)-\psi\left(1/2\right)-\gamma$.
Note that in the flat case the interaction potential,
$e^2/(2\pi)\ln(L)$, tends to $\infty$, when $L\to \infty$, and the
system is in confining phase. Therefore one can conclude that the
geometry of the Poincare upper half space which is characterized
by a constant curvature, gives rise to the screening effect in
three dimensional $QED$.  In the flat case and when the dielectric
constant of vacuum is larger than unity the screening effect (due
to vacuum polarization) may be occurred. So we can assume that the
role of dielectric medium in the flat case is played by the
geometry in our model. To find some relations between geometry and
permittivity see \cite{mol}.
\section{Screening
aspects of Maxwell-Chern-Simons theory on upper half space} In
this part we study the charges potential, when the action
(\ref{2}) is modified by the presence of the Chern-Simons term
\cite{chern}
\begin{equation} \label{18}
S=S_{Max}+S_{SC}=\int_{H^3}\left(-\frac{1}{4}g^{\mu \nu}g^{\lambda
\beta}F_{\mu \lambda}F_{\lambda
\beta}-\frac{s^2}{8\pi}\hat{\epsilon}^{\mu \nu
\rho}A_{\mu}\partial_{\nu}A_{\rho}+J^{\mu}A_{\mu}\right)\sqrt{g}d^3x,
\end{equation}
where $\hat{\epsilon}^{\mu \nu \rho}=1/{\sqrt{g}}\epsilon^{\mu \nu
\rho}$, is the totally antisymmetric Levi-Cevita tensor in the
curved space-time, and $s$ is a numerical constant. The classical
equation of motion of the gauge fields in $H^3$ is
\begin{equation} \label{19}
-\frac{s^2}{4\pi}\epsilon^{\mu \nu
\rho}\partial_{\nu}A_{\rho}+\eta^{\alpha \gamma}\eta^{\mu
\lambda}\partial_{\alpha}\frac{y}{l}F_{\gamma
\lambda}+\sqrt{g}J^{\mu}=0,
\end{equation}
where $\eta^{00}=-\eta^{11}=1$ and $\eta^{ij}=0$ for $i\neq j$.
For static fields, and for the current (\ref{4}), the equations of
motion become
\begin{eqnarray} \label{20}
-\frac{s^2}{4\pi}F_{xy}-\left(\partial_x\frac{y}{l}\partial_x
+\partial_y\frac{y}{l}\partial_y\right)A_0&=&-\sqrt{g}J^0\\
\nonumber
 -\frac{s^2}{4\pi}\partial_yA_0+\partial_y\frac{y}{l}F_{yx}&=&0\\
 \nonumber
\frac{s^2}{4\pi}\partial_xA_0+\partial_x\frac{y}{l}F_{xy}&=&0.
\end{eqnarray}
From the second and the third equations of (\ref{20}), we obtain
$F_{xy}=-ls^2/(4\pi y)A_0$. Putting this expression in the first
equation of (\ref{20}), gives
\begin{equation} \label{21}
-\left(\frac{s^2}{4\pi}\right)^2\frac{l}{y}A_0+\left(\partial_x\frac{y}{l}\partial_x
+\partial_y\frac{y}{l}\partial_y\right)A_0=\sqrt{g}J^0.
\end{equation}
This equation can be solved by construction the Green function
\begin{equation} \label{22}
\left[-\left(\frac{ls^2}{4\pi}\right)^2+y(\partial_xy\partial_x
+\partial_yy\partial_y)\right]G(x,y,x',y')=ly\delta(x,x')\delta(y,y').
\end{equation}
Inserting the Fourier expansion (\ref{7}), in (\ref{22}) we obtain
\begin{equation} \label{23}
G_k(y,y')=-lK_{M}(ky_{>})I_{M}(ky_{<}),
\end{equation}
where $M:=ls^2/(4\pi)$. Therefore the Green function is
\begin{equation} \label{24}
G(x,y;x',y')=\frac{-1}{2\pi}
\sqrt{\frac{l}{y}}\sqrt{\frac{l}{y'}}Q_{M-\frac{1}{2}}\left(\cosh(\frac{L}{l})\right).
\end{equation}
The electrostatic potential of charges is then
\begin{equation} \label{25}
U_{int.}=\frac{-e^2}{2\pi}
\sqrt{\frac{l}{y}}\sqrt{\frac{l}{y'}}Q_{M-\frac{1}{2}}\left(\cosh(\frac{L}{l})\right).
\end{equation}
In the flat space-time limit, $g_{\mu\nu}=\eta_{\mu \nu}$, $l\to
\infty$, the above equation becomes
\begin{equation} \label{26}
U_{int.}=-\frac{e^2}{2\pi}K_0\left(\frac{s^2L}{4\pi}\right),
\end{equation}
which is the same as what was obtained in \cite{ab}.

By considering (\ref{h}), one can show that by increasing $L$, the
interaction potential (\ref{25}), tends faster than (\ref{16})
toward zero. For larger value of $s$, this rate become faster.
This shows that the charges are screened in the presence of the
Chern-Simons term. To see how this effect has been occurred, let
us consider the partition function of $QED_3$ on curved space
-time
\begin{eqnarray}\label{27}
Z=&\int& D
\bar{\psi}D\psi DA_{\mu}\delta(\partial_{\mu}A^{\mu})exp(iS_{Max.})\\
\nonumber &&\exp\left[i\int(\bar{\psi}\gamma
^\mu(i\nabla_{\mu}-qA_{\mu})\psi-m\bar{\psi}\psi)\sqrt{g}
d^3x\right],
\end{eqnarray}
where $q$ (not to be confused with $e$, which is the charge of
external test charges) and $m$ is the charge and the mass of
dynamical fermions repectively, $\nabla$ is the covariant
derivative including the spin-connections and $\gamma^{\mu}$ are
Dirac matrices in curved space-time.

Under the conformal transformation
\begin{eqnarray} \label{28}
\psi\to \lambda&=&\Omega^{-1}\psi \\ \nonumber \bar{\psi}\to
\bar{\lambda}&=&\Omega^{-1}\bar{\psi} \\ \nonumber g_{\mu \nu}\to
\bar{g_{\mu \nu}}&=&\Omega^2g_{\mu \nu},
\end{eqnarray}
the massless part of the fermionic action remains unchanged
\cite{davis}. Therefore by choosing $\Omega=y/l$, in the absence
of conformal anomaly in three dimensions, the partition function
reduces to \cite{eb}
\begin{eqnarray} \label{29}
Z=&\int& D
\bar{\psi}D\psi DA_{\mu}\delta(\partial_{\mu}A^{\mu})exp(iS_{Max.})\\
\nonumber &&\exp \left[i\int(\bar{\psi}\gamma
^a(i\partial_a-qA_a)\psi-\frac{ml}{y}\bar{\psi}\psi) d^3x \right].
\end{eqnarray}
The fermionic part of the action is the same as the flat case with
a position dependent mass term. By integrating out the fermionic
degrees of freedom, up to the order $O(1/m)$ \cite{bos}, we obtain
a $\it{quadratic}$ effective action for the gauge fields
containing the effects of fermionic loop, in the same way as in
the flat case \cite{eb},\cite{ab}:
\begin{equation}\label{30}
S=S_{Max.}+S_{eff.}=S_{Max.}+\int_{H^3}\left(-\frac{q^2}{8\pi}{\epsilon}^{\mu
\nu \rho}A_{\mu}\partial_{\nu}A_{\rho}+O(\frac{1}{m})\right)d^3x.
\end{equation}
So the quantum effects, due to vacuum polarization, can be seen
only through the charge of dynamical fermions $q$ in (\ref{30}).
Therefore to reobtain classical results, i.e. ignoring the vacuum
polarization, it is enough to set $q=0$, in (\ref{30}). By taking
$s=q$, (\ref{30}) reduces to (\ref{18}). Therefore the screening
effect of the model (\ref{18}) is due to the presence of dynamical
fermions. Setting $q=s=0$ results $M=0$, and the $QED$ potential
(\ref{25}), reduces to the potential of test charges in the
absence of vacuum polarization, i.e. (\ref{16}).  The difference
with respect to the flat case is that, in $H^3$ and in the absence
of vacuum polarization, the system is not in confining phase and
dynamical fermions only reduce the absolute value of the
potential.

The effect of geometry on interaction potential of charges can be
extracted from the solutions of the eqs.(\ref{6}) and (\ref{22}),
which contain the metric tensor components as well as their
derivatives. This is due to the presence of the metric dependent
coefficient in Maxwell kinetic term, which has the same role as an
{\it inhomogeneous} dielectric function. Therefore the general
discussion about confinement in the introduction which was based
only on the value of dielectric {\it constant}, is not applicable.
Besides local geometrical effects, boundary conditions considered
for the gauge fields has also direct influence on the charges
interaction potential \cite{sad}, \cite{sah}. For a general
metric, Green functions corresponding to eqs.(\ref{6}) and
(\ref{22}), cannot be obtained exactly. So approximative methods
such as WKB approximation for slowly varying metrics, or adiabatic
expansion of Green function, Taylor expansion of the metric and so
on may be used \cite{sad}. Note that for a metric which is not
conformally flat, the method used in this paper to obtain the
determinant of Dirac operator, is no more valid.

{\bf Acknowledgement:} The author would like to thank University
of Tehran for supporting him under the grant provided by its
Research Council.

\end{document}